# Asymmetry of modal profiles of dipole-exchange spin waves in thin high-magnetic moment metallic ferromagnetic films

*M. Kostylev*

*Abstract*: The asymmetry of the modal profiles for dipole-exchange spin waves propagating in in-plane magnetized ferromagnetic films at a right angle to the applied magnetic field has been investigated theoretically. It was found that in the large-magnetic moment ferromagnetic metallic films with typical thicknesses 10-60 nm the fundamental mode of the spectrum is localized at the surface opposite to the surface of localization of the exchange-free Damon-Eshbach surface wave. This "anomalous" localization of the wave does not affect the non-reciprocity of spin wave excitation by microstrip and coplanar transducers but may be detected in other types of experiments.

The magnetostatic spin waves (SW) propagating in *in-plane* magnetized magnetic films and planar nanostructures with nanometer-range thicknesses have attracted a lot of attention in the recent years due to their potential applications in magnonic devices [1].

Given the largest group velocity of the spin wave which propagates at 90 degree to the applied field ($\mathbf{k} \perp \mathbf{H}$) – the so called Surface or Damon-Eshbach (DE) Wave – has attracted the most of attention [2-8]. In 80ies-90ies this type of spin waves was extensively studied for thick (5micron+) low-magnetic moment yttrium-iron oxide (YIG) epitaxial films. The typical thickness of a YIG film is much larger than $\pi\sqrt{\alpha}$, where $\alpha = 3.1 \cdot 10^{-12}$ cm$^2$ is the exchange constant length for this material, therefore the out-of-plane exchange contribution to the magnetic energy is negligible for these films. As a result, the exchange-free description of the wave spectrum is appropriate [9]. In the framework of this approach it has been found that the Damon-Eshbach wave has a number of properties which are unique with respect to the other types of SW. These are: (i) the spectrum of these waves is single-mode; (ii) the distribution of dynamic magnetization across the film thickness $\mathbf{m}(x)$ ("the modal profile") is given by an asymmetric exponential function exp(-$kx$) ("surface wave") such that the wave propagating in one direction with respect to the direction of the applied field $\mathbf{H}$ is localised at one of the film surface (say, $x=0$ for $H>0$ and $k>0$), whereas the wave travelling in the opposite direction ($k<0$) is localised at the other film surface $x=L$ (here $L$ is the film thickness); and (iii) there is a strong difference in amplitudes of excitation of DE waves by microstrip transducers [10] and a similar property of strong Stokes-anti-Stokes asymmetry of Brillouin light scattering (BLS) peaks [11-13] (for a very simple descriptions see appendices to [14] and [15]) for the waves travelling in two opposite directions with respect to the direction of the applied field ("amplitude non-reciprocity").

Note that we intentionally separated (ii) and (iii) in two different properties, since, as has been shown e.g. in [16], the amplitude non-reciprocity (iii) is not a direct consequence of the modal-profile non-reciprocity (ii) but originates from the important peculiarity of polarisation of dynamic magnetization for this wave. For instance, for the vanishing in-plane wave number ($k=0$) there is no (ii) at all but (iii) is large.

On the other hand, both (ii) and (iii) have the same origin: the DE wave is the only standard type of spin waves in magnetic films for which not one, but two components of the dynamic demagnetizing (dipole) field $\mathbf{h}_d$ give contributions to the wave energy (in the linear case, of course). The difference is, however, in the role of the anti-diagonal components of the magnetostatic Green's function, which describe the dynamic demagnetizing field of precessing magnetization. These components give the contribution to the total amplitude of the in-plane component of the field $h_{dz}$ originating from the out-of-plane component of dynamic magnetization $m_x$ and vice versa ($h_{dx}$ induced by $m_z$). For the description of the amplitude non-reciprocity these components are irrelevant and may be neglected [16-18], but they play the dominant role in the formation of the asymmetric surface-type modal profile for this wave.

Here one has to note that the commonly used formulation of the property (i) in the form as above is not completely correct: formally, there are also an infinite number of volume modes which co-exist with the DE wave for $\mathbf{k} \perp \mathbf{H}$. However, the spectrum of these modes is degenerate and they are dispersionless: the frequency $\gamma[H(H+4\pi M_s)]^{1/2}$ is the same for any wave number $k$ and any mode number $n$ (Here $\gamma$ is the gyromagnetic coefficient and $M_s$ is the film saturation magnetization.) Fig. 3 in the classical paper by Damon and Eshbach [9] explains this fact quite clearly.

For thin magnetic films the contribution of the out-of-plane exchange interaction to the total effective field $\alpha 4\pi M_s(n\pi/L)^2$, becomes comparable to the contribution of the dipole field for small $n$. This lifts the degeneracy of this spectrum: the modes become up-shifted in frequency with a frequency shift scaling roughly as $n^2$. They still remain largely dispersionless until the longitudinal exchange $\alpha 4\pi M_s k_z^2$ kicks in for large in-plane wave numbers $k_z$. (see .e.g Fig.3(d) in [19]).

For the low-magnetic moment epitaxial YIG films with thicknesses 1 to 5 micron the frequency up-shift for the lowest volume modes is usually not large. This results in repulsion of dispersion branches for the *fundamental* mode (the DE wave) and for a respective volume mode and hybridisation of modal profiles in the frequency range where these modes repulse each other (see e.g. Fig. 1 in [20] and Figs.2 and 3 in [30]). As a result, the Damon-Eschbach exchange-free model becomes invalid and the spectrum requires a more rigorous dipole-exchange treatment [20-31].

For the high-magnetic moment metallic films which typically have thicknesses in the range 1nm to 40nm the situation is quite different. The first anti-symmetric volume mode $n=1$ is usually located above the high-frequency limit of existence of the exchange-free DE mode $\gamma(H+2\pi M_s)$. (The volume waves often termed as standing spin wave modes (SSW) in the metallic-film community.) This significant up-shift effectively makes the fundamental mode $n=0$ exchange-free again and the Damon-Eshbach model is applicable to its description again and is largely used in the community for the moderate wave numbers for which the longitudinal exchange interaction can be neglected (see e.g. [4,32,33]).

In the following we will show that although this "effective exchange-free" approach is well suited for description of the dispersion of the fundamental mode it does not provide proper description of the modal non-reciprocity. We show that in the picture of mode hybridization the surface character of the fundamental mode originates from the dipole coupling of the spectrum of the volume (i.e. SSW) modes with the $n=0$ mode via the anti-diagonal component of the magneto-static Green's function [34]. When the degeneracy of this spectrum is lifted for $\alpha \neq 0$, the mode coupling becomes strongly dependent on the frequency position of the 1st SSW with respect to the frequency of the fundamental mode for a particular wave number. If the exchange field is small (including vanishing) and, consequently, the 1st SSW is located below the fundamental mode the fundamental mode is localised at the same surface as the "true" DE wave $\alpha=0$. However, if the 1st SSW is strongly pushed upwards by a small film thickness and large saturation magnetization the $n=0$ mode is localised at the surface opposite to one for the true DE wave.

Note that this property has no relevance for the experiments on excitation of spin waves by strip-line antennas [32,36], since the results are mostly dependent on the amplitude non-reciprocity which remains the same, because the fundamental mode polarisation is almost the same as in the exchange-free case. However, it is of importance for the Spintronics-related experiments on spin wave Doppler shift [37] since this effect may give an important unwanted contribution to the total frequency shift when a dc current flows through the film co-linear with the SW wave vector [38].

In the next section we present an approximate analytical theory of this effect and the results of numerical calculations supporting this analytical theory. Conclusions are contained in Section III.

## II Theory.

In order to solve the problem we will follow the approach from [31]. For the self-consistency of the paper we repeat it here. Consider the frame of reference already partly introduced in Section I. The $x$-axis is perpendicular to the surface of a magnetic film. The surface is continuous in both $y$ and $z$ directions. The in-plane axis $z$ coincides with the direction of the positive SW vector $k_z$ of a plane spin wave. The magnetic field $\mathbf{H}$ is applied in the film plane in the positive direction of the $y$-axis.

To describe the magnetization dynamics we will use the linearized Landau-Lifshitz equation

$$\partial \mathbf{m} / \partial t = -\gamma \mathbf{m} \times \mathbf{H} + \mathbf{h}_{eff} \times \mathbf{M}. \quad (1)$$

Here the dynamic magnetization vector $\mathbf{m}$ has only two non-vanishing components $(m_x, m_z)$ which are perpendicular to the static (equilibrium) magnetization vector $\mathbf{M} = M_s \cdot \mathbf{e}_y$, $\mathbf{e}_y$ is the unit vector in the $y$-direction, and the dynamic effective field $\mathbf{h}_{eff}$ has two components: the exchange field $\mathbf{h}_{ex}$ given by

$$\mathbf{h}_{ex} = \alpha (\partial^2 / \partial x^2 + \partial^2 / \partial z^2) \mathbf{m} \quad (2)$$

and the dynamic demagnetizing (dipole) field $\mathbf{h}_d$. We seek the solution of (1) in the form of a plane spin wave

$$\mathbf{m}, \mathbf{h}_{eff} = \mathbf{m}_k, \mathbf{h}_{effk} \exp(i\omega t - k_z z). \quad (3)$$

The dipole field is given by the magnetostatic Green's function in the Fourier space $\mathbf{G}_k(s)$ [31]

$$\mathbf{h}_{dk}(x) = \int_0^L \mathbf{G}_k(x-x') \mathbf{m}_k(x') dx' \equiv \mathbf{G}_k \otimes \mathbf{m}_k. \quad (4)$$

In our frame of reference the components of this function take the form

$$\mathbf{G}_k(s) = \begin{pmatrix} G_{kxx} & G_{kxz} \\ G_{kzx} & G_{kzz} \end{pmatrix} = 4\pi \begin{pmatrix} \delta(s) - G_p(k,s) & iG_q(k,s) \\ iG_q(k,s) & -G_p(k,s) \end{pmatrix}, \quad (5)$$

where $G_p = \dfrac{|k_z|}{2}\exp(-|k_z||s|)$, $G_q = \text{sign}(s)\dfrac{k_z}{2}\exp(-|k_z||s|)$, $\delta(s)$ is Dirac delta function, and sign($s$)=1 for $s>0$ and $-1$ for $s<1$. Note that the only place where the sign of $k_z$ matters is in the pre-factor of the expression for $G_q$. Thus, the whole information about the non-reciprocity of SW is contained in the sign of this pre-factor. We now substitute (2)-(4) into (1) and perform a co-ordinate transformation $m_{xk}=(m_k^{(1)}+m_k^{(2)})/2$, $m_{yk}=(m_k^{(1)}-m_k^{(2)})/(2i)$ and a similar transformation for the components of $\mathbf{h}_{effk}$. In these circular co-ordinates the linearized Landau-Lifshitz equation takes a very simple form

$$\omega \mathbf{m}_k = \begin{pmatrix} -[\omega_H + \omega_M(\alpha\partial^2/\partial x^2 - \alpha k_z^2 + 1/2)]\delta & \omega_M(G_q + G_p - \delta/2) \\ \omega_M(G_q - G_p + \delta/2) & [\omega_H + \omega_M(\alpha\partial^2/\partial x^2 - \alpha k_z^2 + 1/2)]\delta \end{pmatrix} \otimes \mathbf{m}_k \quad (5)$$

where $\delta=\delta(s)$ (the Dirac delta function, as above), $\omega_H = \gamma H$ and $\omega_M = \gamma 4\pi M_s$ and $\mathbf{m}_k$ now has components $(m_k^{(1)}, m_k^{(2)})$.

One sees that the eigenfrequency of spin waves represents an eigenvalue of the integro-differential operator given by the brackets on the right-hand side of (5). Accordingly, the eigenfunctions of the operator represent the modal profiles for the respective spin wave modes.

To obtain the eigenspectrum of spin waves and the modal profiles we solve Eq.(5) numerically. The mesh size is chosen such that the dipole sums resulting from discretization of the magnetostatic Green's function (5) converge well. For instance, to calculate the dispersion of the fundamental mode for a 40nm-thick Permalloy film 200 points across the film thickness are enough in the wave number range of interest $0<k_z<10^5$ rad/cm.

A typical result of the numerical calculations is shown in Figs. 1 and 2.

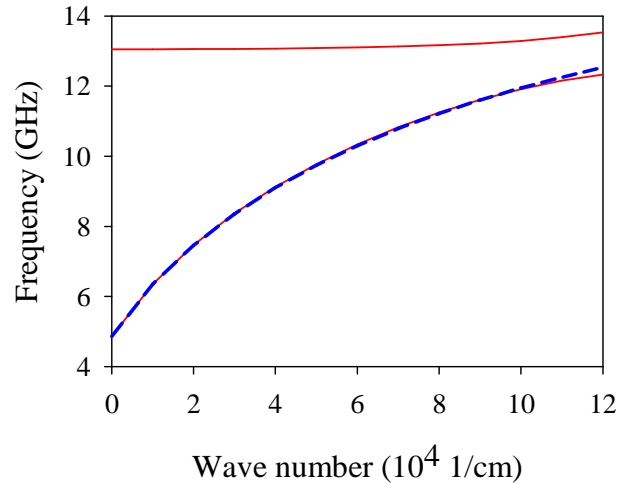

Fig. 1. Dispersion of the exchange-free Damon-Eshbach wave (thick dashed line) and dispersion of the fundamental (lower-frequency) and the 1$^{st}$ exchange (higher-frequency) mode of dipole exchange spin waves propagating at 90 degrees to the applied field in a tangentially magnetized Permalloy film (thin solid lines). Film thickness is 40nm, applied field is 280 Oe. Saturation magnetization $4\pi M_s$=10500 G.

Figure 1 displays the dispersion of the fundamental mode of a 40nm-thick Permalloy film calculated in exchange-free approximation ($\alpha, A$=0) and using a realistic value for the exchange constant for Permalloy $A$=10$^{-6}$ erg/cm. Note that the obtained exchange-free dispersion is in full quantitative agreement with the Damon-Eshbach formula [9] and that in the dipole-exchange case we used the "unpinned surface spins" exchange boundary conditions [39]. One sees that the difference between the two dispersions negligible in the whole wave-number range displayed in the figure except the small range $k$>10$^5$ rad/cm where the fundamental mode approaches the 1$^{st}$ SSW mode.

Figure 2 shows the respective modal profiles calculated for $k$=78000 rad/cm which is outside the range where the fundamental and the 1$^{st}$ SSW modes start to repulse each other. The main observation from the figure is that the exchange-free and the dipole exchange waves are localised at the opposite film surfaces. One also sees that the asymmetry (localisation) of the dipole-exchange mode is stronger that of the exchange-free wave. The location of the surface at which the exchange-free wave is localised is in full agreement with the criterion for the DE wave [40] **k**/|**k**|=**n**$_0$×**M**/|**M**|, where **n**$_0$ is the internal normal to the surface at which the wave is localised. Thus, the exchange interaction results in localisation of the fundamental mode at the surface opposite to the standard surface of localisation of the DE wave. The last statement is the main finding of this work.

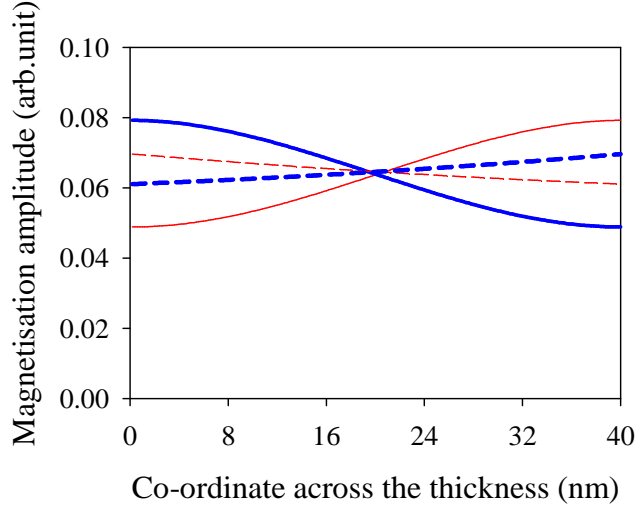

Fig. 2. Modal profiles for the exchange-free Damon Eshbach wave (dashed lines) and for the fundamental mode of the dipole-exchange spin waves (solid lines). Thick lines: $k_z$=−78000 rad/cm, thin lines: $k_z$=+78000 rad/cm. All other parameters as for Fig. 1.

Now we show a simple analytical theory which provides explanation of the observed phenomenon and clearly shows the difference between the true "Damon Eschbach" dispersion for thick films and the "pseudo Damon-Eschbach" behaviour of the fundamental mode of dipole exchange SW in thin metallic magnetic films.

To construct this theory, as in [31] we solve the integro-differential equation (5) with Bubnov-Galerkin method. It consists in expanding the *x*-dependence of $\mathbf{m}_k$ into series using an ortho-normal system of functions satisfying the exchange boundary conditions at both film surfaces. For simplicity we assume the unpinned surface spins boundary conditions ($\partial \mathbf{m}_k / \partial x|_{x=0,L} = 0$) [39]. The following orthogonal series satisfies these conditions

$$\mathbf{m}_k(x) = \mathbf{m}_{k0} + \sqrt{2}\sum_{i=1}^{\infty} \mathbf{m}_{ki} \cos\left(\frac{i\pi}{L}x\right). \quad (6)$$

By projecting this equation on the basis of these orthogonal functions we obtain an infinite system of homogeneous algebraic equations given by Eq.(22) in [31].

As shown in [19], the eigenfequencies of dipole-exchange spin wave can be calculated by numerically solving the eigenvalue problem for the matrix of coefficients of this system. Similarly, the eigenvectors of this matrix give the amplitudes of the series expansion $\mathbf{m}_{ki}$ of the wave modal profiles $\mathbf{m}_k(x)$ (see the discussion in [19] regarding Fig. 4 in that paper). In contrast to this previous study here we will solve the eigenvalue/eigenvector analytically approximately to find the modal profiles of the DE-wave in the dipole-exchange regime.

For this purpose we assume that the film is metallic magnetic and has a thickness typical for those samples such that the second SSW has the frequency lying well above the upper boundary of existence of the DE wave. In these conditions the DE wave may only efficiently couple to the 1st SSW which, depending on the film thickness, may be located either within the DE-wave frequency band (and both may repulse each other) or at any frequency distance above the upper frequency edge of existence of the DE wave. Under this assumption we may truncate the series (6) after *i*=1. The respective vector-matrix equation takes the form

$$\omega \tilde{\mathbf{m}}_k = \begin{pmatrix} \eta_0 & \theta \\ -\theta & \eta_1 \end{pmatrix} \tilde{\mathbf{m}}_k \equiv C_k \tilde{\mathbf{m}}_k, \quad (7)$$

where $\tilde{\mathbf{m}}_k$ is a 4-element column vector ($m_{k0}^{(1)}$, $m_{k0}^{(2)}$, $m_{k1}^{(1)}$, $m_{k1}^{(2)}$) and the 2x2 blocks $\eta_{0(1)}$ of the 4x4 matrix $C_k$ are as follows.

$$\eta_{0(1)} = \begin{pmatrix} A_{0(1)} & B_{0(1)} \\ -B_{0(1)} & -A_{0(1)} \end{pmatrix}, \quad (8)$$

where

$$A_0 = -\omega_H - \omega_M/2, \; A_1 = A_0 - \omega_M \alpha (\pi/L)^2, \; B_1 = -\omega_M/2, \; B_0 = B_1 - \omega_M P_{00}(k) \quad (9)$$

and

$$P_{ij}(k_z) = 2\int_o^L \cos(i\pi x/L)dx \int_0^L G_p(x-x')\cos(j\pi x'/L)dx' \quad (10)$$

is the matrix element originating from projection of the $G_p$ component of the magnetostatic Green's function (4) on the basis of the functions (6). The $P_{00}$ element has an exceptionally simple form: $1-[1-\exp(-|k_z|L)]/(|k|_z L)$. In order to simplify the expression, while writing down (9), in $A_0$ and $A_1$ we neglected the terms $\omega_M \alpha(k_z)^2$ responsible for the longitudinal exchange. This is a valid approximation for the most of the wave number range $\alpha(k_z)^2 \ll P_{00}(k_z)$ excited by the microwave transducers and accessible with BLS. We also neglected the term $\omega_M P_{11}(k_z)$ in $B_1$, because $P_{11} \ll P_{00}$ for the same small-wave-number range and because from the exact numerical solution we know that SSW are dispersionless in this wave number range (see the discussion in the introduction).

The matrix $\theta$ is anti-diagonal

$$\theta = \omega_M Q \begin{pmatrix} 0 & 1 \\ 1 & 0 \end{pmatrix}, \quad (10)$$

where $Q \equiv Q_{01}(k_z) = \sqrt{2} \int_o^L \cos(0 \pi x / L) dx \int_0^L G_q(x - x') \cos(1 \pi x' / L) dx'$. In the closed form it is given by Eq.(25) in [31].

As shown in [41] a good approximation to the mode dispersion is given by the eigenvalues of the diagonal blocks of the matrix $C_k$. The eigenvalues of $\eta_0$ give the dispersion of the fundamental mode $\omega_{k,n=0}$ and the eigenvalues of $\eta_1$ the dispersion of the 1$^{st}$ SSW $\omega_{k,n=1}$. These approximate relations are valid far away from the points of intersection of the dispersion branches they describe. Near the intersection points where the branches of the exact spectrum repulse each other and the mode hybridize the mode coupling should be included. As follows from (7) in the present case the $\theta$ blocks are responsible for the coupling between the diagonal blocks.

The effect of this coupling can be obtained in the framework of the theory of perturbations for the matrix eigenvalues. We will consider the matrix $C_k^{(0)} = \begin{pmatrix} \eta_0 & 0 \\ 0 & \eta_1 \end{pmatrix}$ as the unperturbed matrix and the matrix $C_k^{(1)} = \begin{pmatrix} 0 & \theta \\ -\theta & 0 \end{pmatrix}$ as the matrix of perturbation. The latter is appropriate, given $|Q| \ll P_{00}$.

The perturbation of a matrix eigenvector is given by the admixture of the other unperturbed eigenvectors to the eigenvector. In our case of coupling of just two modes the expression for the perturbed eigenvector for the fundamental mode is given by

$$\tilde{\mathbf{m}}_{k,n=0} = \tilde{\mathbf{m}}_{k,n=0}^{(0)} + c_k \tilde{\mathbf{m}}_{k,n=1}^{(0)}, \quad (11)$$

where the amplitude $c_k$ of admixture of the 1$^{st}$ SSW is given by

$$c_k = \frac{\boldsymbol{\mu}_{k,n=1}^{(0)} C_k^{(1)} \tilde{\mathbf{m}}_{k,n=0}^{(0)}}{\omega_{k,n=0} - \omega_{k,n=1}}, \quad (12)$$

$\tilde{\mathbf{m}}^{(0)}_{k,n=0}$ and $\tilde{\mathbf{m}}^{(0)}_{k,n=1}$ are the unperturbed right-hand (column) eigenvectors for the fundamental mode and for the 1$^{st}$ SSW respectively and $\boldsymbol{\mu}^{(0)}_{k,n=1}$ is the left-hand (row) eigenvector of $C^{(0)}_k$ corresponding to the unperturbed 1$^{st}$ SSW [42].

The solution of the eigenvalue/eigenvector problem for a matrix in the form (8) is known. It is given by Bogolyubov transformation [43]. The eigenvalues are

$$\omega_{k,n} = \pm\sqrt{A_n^2 - B_n^2}. \quad (13)$$

The right-hand eigenvectors $\tilde{\mathbf{m}}^{(0)}_{k,n}$ are $(u_n, v_n)$ and $(v_n, u_n)$ for $-|\omega_{k,n}|$ and $+|\omega_{k,n}|$ eigenvalues respectively, where $u_n = \sqrt{(|\omega_{k,n}| - A_n)/(2|\omega_{k,n}|)}$ and $v_n = -\sqrt{(-|\omega_{k,n}| - A_n)/(2|\omega_{k,n}|)}$. (Note that the negative sign in front of the square root in $v_n$ is because in our case $B_n < 0$.) The respective left-hand eigenvectors are $(u_n, -v_n)$ and $(v_n, -u_n)$.

Thus, the amplitude $c_k$ is given by a very simple expression

$$c_k = \frac{1}{\omega_{k,n=0} - \omega_{k,n=1}} \begin{pmatrix} 0 & 0 & -v_2 & u_2 \end{pmatrix} C_k \begin{pmatrix} v_1 \\ u_1 \\ 0 \\ 0 \end{pmatrix}, \quad (14)$$

or, in the closed form,

$$c_k = \frac{\omega_M Q}{\omega_{k,n=0} - \omega_{k,n=1}}(u_1 v_2 - u_2 v_1). \quad (15).$$

Let us examine the sign of $c_k$ first as a function of the wave number $k_z$. The quantities $\omega_{k,n=1}$, $u_2$ and $v_2$ do not depend on $k_z$. Furthermore, $u_{1(2)} > 0$, $v_{1(2)} < 0$, $u_1 > |v_1|$ and the ratio $|v_1|/u_1$ decreases with an increase in $k_z$. This follows from the decrease in the DE wave ellipticity with $k_z$. Thus, the expression in the brackets on r.h.s. of (15) represents a difference of two positive quantities. So, it may change sign as a function of $k_z$. However, the dependence of $u_1$ and $v_1$ on $k_z$ is slower than the dependence of $\omega_{k,n=0}$ on $k_z$, since the eigenvectors of a matrix are more stable to variation of its parameters than the eigenvalues (the same rule (12) applies).

Thus, one sees that in the first place the sign of $c_k$ is given by the sign of the difference in the denominator of (15): for the set of parameters for Fig. 2 if the frequency of the unperturbed fundamental mode is higher than the frequency of the unperturbed 1$^{st}$ SSW $c_k$ is negative. If it is lower, then $c_k$ is positive in a wide range of wave numbers from both sides of the point of the intersection of the unperturbed branches.

The physical meaning of $c_k$ follows from the expression for the perturbed modal profile: $\mathbf{m}_{k,n=0}(x) = \tilde{\mathbf{m}}^{(0)}_{k,n=0} + c_k \sqrt{2} \tilde{\mathbf{m}}^{(0)}_{k,n=1} \cos\left(\frac{\pi}{L}x\right)$. One sees that the perturbed profile is a combination of a uniform function and a cosine function. The contribution of the cosine function to the total profile is anti-symmetric across the film thickness. Thus, the maximum of the total perturbed profile $\left|m^{(2)}_{k,n=0}(x)\right|$ is located at one of the film surfaces and the location of this surface is given by the sign of $u_2/u_1$.

Consider ther wave $+\left|\omega_{k,n}\right|$ propagating in the positive direction of the axis $z$. For the wave localised at the same surface as the exchange-free DE wave this ratio should be positive. Figure 3(b) displays this ratio as a function of the value of the exchange constant calculated by using Eqs.(11) and (15). Figure 3(a) shows the respective perturbed eigenfrequencies calculated as the eigenvalues of $C_k$. The value of the SW wave number for this calculation $k_z$=78000 1/cm. Since in our approximate model we neglect the exchange contribution to the frequency of the fundamental mode, the fundamental mode frequency is given by the horizontal straight line $\omega/(2\pi)$=10.710GHz in Fig.3(a). The curved sections of the lines are the perturbed frequency of the 1$^{st}$ SSW. The area where the straight line and the curved line repulse each other is the frequency range of the mode hybridisation.

The main observation here is that when the exchange constant is small (including vanishing) the sign of this ratio is negative, i.e. the same as for the exchange-free Damon-Eschbach wave. The frequency of the fundamental mode is smaller than the frequency of the 1$^{st}$ SSW in this range. Above the point of intersection of the two unperturbed dispersions $\omega_{k,n=1} > \omega_{k,n=0}$ the sign of this ratio is positive, which implies that the fundamental mode is localized at the opposite surface now. This result is in full agreement with the exact solution from Fig. 2. Thus, this calculation confirms our conclusion above that the localisation is given by the sign of the frequency difference in the denominator of Eq.(15).

Importantly, the sign remains the same far away from the hybridisation point. This implies that in the case of the thin metallic films (Fig. 1) for which the 1$^{st}$ exchange mode has the frequency usually (well) above the frequency of the fundamental mode the localisation of the fundamental mode of the dipole-exchange spectrum at the surface which is opposite to one at which the exchange free DE wave is localized is the effect of the dipole coupling of the fundamental mode to the nearest exchange branch. Rigorous numerical calculation using Eq.(5) shows that for a given wave number the extent of asymmetry of the mode profile (mode localisation) $(m_k^{(2)}(x=L)-m_k^{(2)}(x=0))/(m_k^{(2)}(x=L)+m_k^{(2)}(x=0))$ depends on the position of the frequency with respect to the hybridization point which can be changed by varying the value of the exchange constant. In particular, if the 1$^{st}$ SSW mode is above the fundamental mode, the modal profile asymmetry (localisation) reduces with increase in the frequency of the 1$^{st}$ SSW i.e. with frequency separation of the two modes. This is in full agreement with Fig. 3(b).

Also in agreement with Fig. 3(b) the localisation remains non-vanishing for $\alpha$=0 and $k\neq 0$. In this particular case it scales as $\exp(-k_zL)$ which is in full agreement with DE theory. This implies that in the framework of the coupling model (7) the localisation of the exchange-free wave at one of the surfaces is due to its coupling to the degenerate spectrum $\omega(k)=\gamma[H(H+4\pi M_s)]^{1/2}$ of the dispersionless bulk waves (see the discussion in the introduction). Importantly, the coupling does not decrease with an increase in the frequency of the DE mode (i.e. the mode localization grows with

$k_z$), possibly because this degenerate spectrum contains an infinite number of modes, and because $|Q_{ij}(k_z)|$ increases with $k_z$. (An example of $Q_{ij}(k_z)$ dependence is shown in Fig. (4).)

Recall that $Q_{ij}(k_z)$ originates from the anti-diagonal elements $G_q$ of the Green's function. It vanishes when both $i$ and $j$ are odd or even. In this way it is responsible for coupling of modes of different symmetries (even to odd). In particular, it is responsible for the admixture of higher anti-symmetric terms $j=1,3,5…$ of (6) to the uniform $j=0$ term in the case of the modal profile for the fundamental mode. This explains the role of the anti-diagonal component of the Green's function in the formation of the surface character of the Damon-Eshbach wave. (Recall that this wave is the only standard type of SW for which $G_q$ gives contribution to the total torque acting on the dynamic magnetization.)

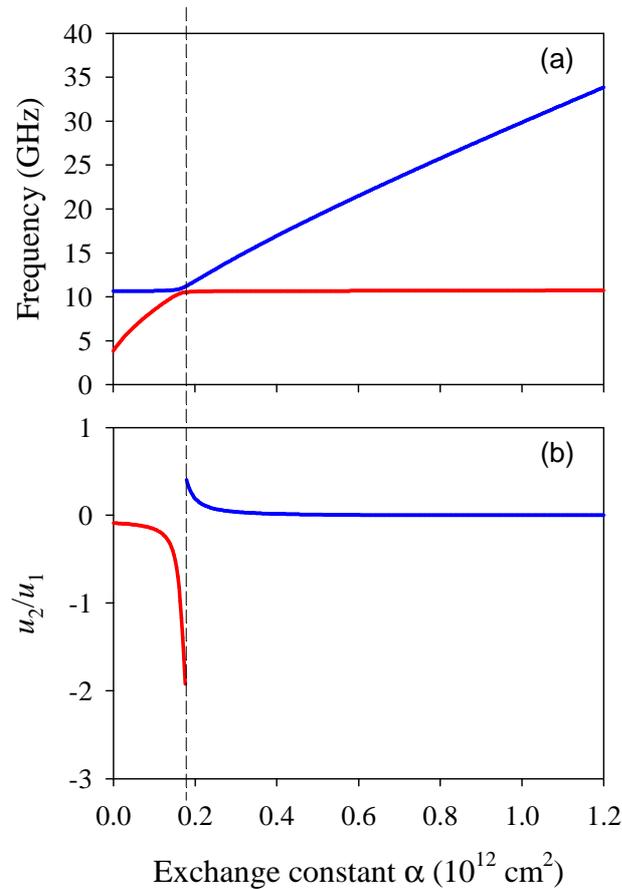

Fig. 3. (a) Dependence of the frequency of spin waves on the value of the exchange constant calculated by solving Eq.(7) numerically. Wave number is 78000 1/cm and all other parameters are as in Fig. 1. (b) The ratio of the Fourier amplitudes $u_2/u_1$ for the same set of parameters.

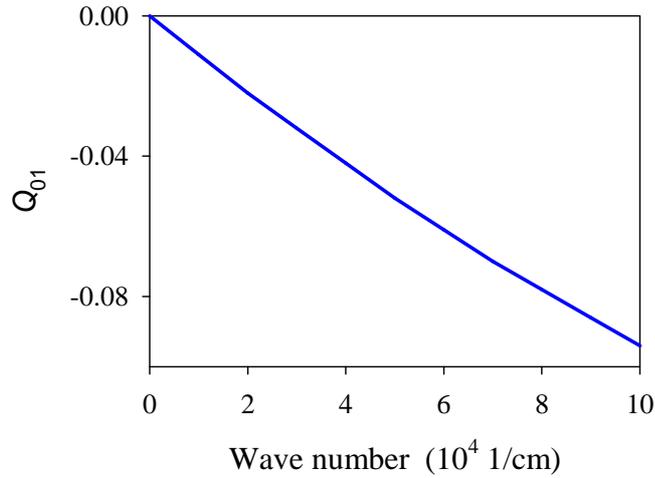

Fig. 4. Matrix element $Q_{01}$ as a function of spin-wave wave number calculated for the parameters of Fig. 1.

### III Conclusion

In this work we have investigated asymmetry of the modal profiles for dipole-exchange spin waves propagating in in-plane magnetized ferromagnetic films at a right angle to the applied magnetic field. We found that in the large-magnetic moment ferromagnetic metallic films such as made from Permalloy, iron or similar with typical thicknesses 10-60 nm the fundamental mode of the spectrum is typically localized at the surface opposite to the surface of localization of the exchange-free Damon-Eshbach surface wave. The latter theory is a good model for the low-magnetic moment 1micron+ yttrium iron garnet ferrite films but, as shown here, fails to adequately describe the mode profiles for the dipole exchange waves in thin high-magnetic-moment ferromagnetic films. We also explained the mode asymmetry as originating from the anti-diagonal component $G_q$ of the tensor Green's function of the magnetostatic (dipole) field. This type of waves is the only standard type of SW for which $G_q$ contributes to the total torque acting on the magnetization vector. In the framework of this approach this explains why only this standard type of SW is characterised by an asymmetry of the modal profile across the film thickness.

**Acknowledgement:** We would like to thank M. Bailleul and B. A. Kalinikos for stimulating discussions leading to understanding of this interesting phenomenon. Funding from Australian Research Council is gratefully acknowledged.

References:
1. V. V. Kruglyak, S. O. Demokritov, and D. Grundler, *J. Phys. D: Appl. Phys.* **43**, 264001 (2010).
2. V. E. Demidov, S. Urazhdin, and S. O. Demokritov, *Appl. Phys. Lett.,* **95**, 262509 (2009).
3. K. Vogt, H. Schultheiss, S. Jain, J. E. Pearson, A. Hoffmann, S. D. Bader, and B. Hillebrands, *Appl. Phys. Lett.*, **101**, 042410 (2012).
4. K.Yu. Guslienko, S. O. Demokritov, B. Hillebrands, and A. N. Slavin, *Phys. Rev. B*, **66**, 132402 (2002).

5. A. V. Chumak, P. Pirro, A. A. Serga, M. P. Kostylev, R. L. Stamps, H. Schultheiss, K. Vogt, S. J. Hermsdoerfer, B. Laegel, P. A. Beck, and B. Hillebrands, *Appl. Phys. Lett*. **95**, 262508 (2009).
6. R. Zivieri, S. Tacchi, F. Montoncello, L. Giovannini, F. Nizzoli, M. Madami,2 G. Gubbiotti, G. Carlotti, S. Neusser, G. Duerr, and D. Grundler, *Phys. Rev. B*, **85**, 012403 (2012).
7. M. Bailleul, D. Olligs, and Claude Fermon, *Appl. Phys. Lett*., **83**, 972 (2003).
8. A. A. Stashkevich, P. Djemia, Y. K. Fetisov, N. Biziere, C. Fermon, *J. Appl. Phys*. **102**, 103905 (2007).
9. R.W. Damon, J.R.Eshbach, *J. Phys. Chem. Solids* **19**, 308 (1961).
10. P. R. Emtage, *J. Appl. Phys*. **49**, 4475 (1978).
11. M. G. Cottam, D. J. Lockwood, "Light scattering in magnetic solids", Wiley, 1986.
12. B. Hillebrands, in: *Light Scattering in Solids VII*, M. Cardona, G. Güntherodt (Eds.), *Topics in Applied Physics* **75**, Springer Verlag, Heidelberg, 174 (1999).
13. T.S. Rahman, D.L.Mills, R.E. Camley, *Phys. Rev. B* **23,** 1226 (1981).
14. A. A. Stashkevich, Y. Roussigné, P. Djemia, S. M. Chérif, P. R. Evans, A.P. Murphy, W. R. Hendren, R. Atkinson, R. J. Pollard, A. V. Zayats, G. Chaboussant, and F. Ott, *Phys. Rev. B* **80**, 144406 (2009).
15. M. Kostylev, A. A. Stashkevich, A. O. Adeyeye, C. Shakespeare, N. Kostylev, N. Ross, K. Kennewell, R. Magaraggia, Y. Roussigné, and R. L. Stamps, *J.Appl. Phys*. **108**, 103914 (2010).
16. T. Schneider, A. A. Serga, T. Neumann, B. Hillebrands, and M. P. Kostylev, *Phys. Rev. B* **77**, 214411 (2008).
17. V. F. Dmitriev and B. A. Kalinikos, *Sov. J. Phys.* **31**, 875 (1988).
18. B. A. Kalinikos, *IEE Proceedings S-H Microw, Anten. Prop*. **127**, 4 (1980).
19. M. P. Kostylev, B. A. Kalinikos, H. Dötsch, *J. Magn. Magn. Mat*. **145**, 93 (1995).
20. R. E. De Wames and T. Wolfram, *J. Appl. Phys*. **41**, 987 (1970).
21. B. N. Filippov, *Phys. Met. Metall*. **32** (1971).
22. B. N. Filippov and I. G. Titjakov, *Phys. Met. Metall*. **35** (1973).
23. L. V. Mikhailovskaya and R. G. Khlebopros, *Sov. Phys. – Solid State* **16**, 46 (1974).
24. Yu. V.Gulyaev, P. E. Zil'berman and A. V. Lugovskoi, *Sov. Phys. – Solid State* **23**, 660 (1981).
25. A. V. Lugovskoi and P. E. Zil'berman, *Sov. Phys. – Solid State* **24**, 259 (1982).
26. M. Sparks, *Phys. Pev. Lett*. **24**, 1178 (1970).
27. M.Sparks, *Phys. Rev. B* **1** 3831 (1970).
28. R. L. Stamps and B. Hillebrands, *Phys. Rev. B* **43**, 3532 (1991).
29. B. A. Kalinikos, "Dipole-exchange spin-wave spectrum of magnetic films" in "Linear and nonlinear spin waves in magnetic films and Superlattices," M. G. Cottam, editor, World Scientific Publishing Company, Ltd., Singapore, 1994.
30. B.A. Kalinikos, M.P. Kostylev, N. V. Kozhus', A. N. Slavin, *J. Phys. Cond. Mat*. **2**, 9861 (1990).
31. B. A. Kalinikos, *Sov. J. Phys* **24**, 718 (1981).
32. V. Vlaminck and M. Bailleul, *J. Phys. Rev. B* **81**, 014425 (2010).
33. F Zighem, Y Roussigne, S-M Cherif, and P Moch, *J. Phys.: Condens. Matter* **19** (2007).

34. Note that the eigenvalues of this function are analogous to the demagnetizing factors for an ellipsoid [35], thus the anti-diagonal element of this tensor plays the role of the anti-diagonal element of the tensor of the demagnetizing factors.
35. G. Gubbiotti, S. Tacchi, M. Madami, G. Carlotti, A. O. Adeyeye and M. Kostylev, *J. Phys. D: Appl. Phys*. **43**, 264003 (2010).
36. E. Demidov, M. P. Kostylev, K. Rott, P. Krzysteczko, G. Reiss, and S. O. Demokritov, *Appl. Phys. Lett*. **95**, 112509 (2009).
37. V. Vlaminck and M. Bailleuil, *Science* **322**, 410 (2008).
38. M.Haidar, M.Bailleul, and M.Kostylev, unpublished.
39. C. T. Rado and T. R. Weertman, *J. Phys. Chem. Solids* **2**, 315 (1959).
40. A.G.Gurevich and G.A.Melkov, "*Magnetisation oscillations and waves*", Sect. 6.2, CRC Press.
41. B.A.Kalinikos and A.N.Slavin, *J.Phys.C: Solid State Phys*. **19**, 7013 (1986).
42. We have to distinguish between the right-hand and the left-hand eigenvectors because $C_k^{(0)}$ is not symmetric. Consequently, its right-hand eigenvectors are not orthogonal. To construct an ortho-normal basis and we then have to use the bi-orthogonality property of the left-hand and the right-hand eigenvectors $\boldsymbol{\mu}_{k,i}^{(0)} \cdot \tilde{\mathbf{m}}_{k,j}^{(0)} = \delta_{i,j}$. (Here *i*,*j* are the eigenvectors corresponding to the *i*-th and *j*-th eigenvalue respectively.) A more detailed explanation of this approach can be found in Ref. [19].
43. V.S.L'vov, "*Wave Turbulence under Parametric Excitations. Applications to Magnetic*", Springer-Verlag, 1994.